\pdfoutput=1

\documentclass[10pt,conference]{IEEEtran}

\usepackage{amsmath,amssymb,amsfonts}
\usepackage{algorithmic}
\usepackage{graphicx}
\usepackage{textcomp}
\usepackage{stfloats}
\usepackage{adjustbox}
\usepackage{diagbox}
\usepackage{float}

\def\BibTeX{{\rm B\kern-.05em{\sc i\kern-.025em b}\kern-.08em
    T\kern-.1667em\lower.7ex\hbox{E}\kern-.125emX}}

\usepackage{url}
\usepackage{array}
\usepackage[table]{xcolor}
\usepackage{float}
\usepackage{listings}
\usepackage{graphicx}
\usepackage{multicol}
\usepackage{booktabs}
\lstdefinestyle{base}{
  language=Java,
  emptylines=1,
  breaklines=true,
  basicstyle=\scriptsize,
  moredelim=**[is][\color{red}]{@}{@},
  floatplacement=H
}
\usepackage{comment}
\usepackage{fancyhdr}
\usepackage[yyyymmdd,hhmmss]{datetime}

\begin{document}

\title{Recover and RELAX: Concern-Oriented Software Architecture Recovery for Systems Development and Maintenance}

\author{\IEEEauthorblockN{Daniel Link\IEEEauthorrefmark{1}, Pooyan Behnamghader\IEEEauthorrefmark{2}, Ramin Moazeni\IEEEauthorrefmark{3} and Barry Boehm\IEEEauthorrefmark{4}} \IEEEauthorblockA{\IEEEauthorrefmark{1}\IEEEauthorrefmark{2}\IEEEauthorrefmark{4}Center for Systems and Software Engineering\\
University of Southern California, Los Angeles, USA\\
\IEEEauthorrefmark{1}Email: dlink@usc.edu
\IEEEauthorrefmark{2}Email: pbehnamg@usc.edu
\IEEEauthorrefmark{4}Email: boehm@usc.edu} \IEEEauthorblockA{\IEEEauthorrefmark{3}Computer Engineering Department\\
Santa Clara University, Santa Clara, USA\\
Email: rmoazzeni@scu.edu}}

\maketitle

\begin{abstract}
The stakeholders of a system are legitimately interested in whether and how its architecture reflects their respective concerns at each point of its development and maintenance processes. Having such knowledge available at all times would enable them to continually adjust their system’s structure at each juncture and reduce the buildup of technical debt that can be hard to reduce once it has persisted over many iterations. 
Unfortunately, software systems often lack reliable and current documentation about their architecture. In order to remedy this situation, researchers have conceived a number of architectural recovery methods, some of them concern-oriented. However, the design choices forming the bases of most existing recovery methods make it so none of them have a complete set of desirable qualities for the purpose stated above. 
Tailoring a recovery to a system is either not possible or only through iterative experiments with numeric parameters. Furthermore, limitations in the scalability of the employed recovery algorithms make it prohibitive to apply the existing techniques to large systems. Finally, since several current recovery methods employ non-deterministic sampling, their inconsistent results do not lend themselves well to tracking a system’s course over several versions, as needed by its stakeholders. 

RELAX (RELiable Architecture EXtraction), a new concern-based recovery method that uses text classification, addresses these issues efficiently (1) by assembling the overall recovery result from smaller, independent parts, (2) basing it on an algorithm with linear time complexity and (3) being tailorable to the recovery of a single system or a sequence thereof through the selection of meaningfully named, semantic topics. An intuitive and informative architectural visualization rounds out RELAX's contributions. RELAX is illustrated on a number of existing open-source systems and compared to other recovery methods.
	
\end{abstract} %
\IEEEpeerreviewmaketitle

\section{Introduction}

A software system can only be maintained to the extent that it is known. Knowing a system includes being aware of its architecture. This awareness avoids technical debt and ensures the system's continued integrity. 

In practice, the knowledge of a system's architecture may have never existed or been degraded over time through phenomena such as missing or poor documentation, personnel changes as well as architectural drift or erosion \cite{Taylor2010}. (The latter two are caused by careless %
or unintentional addition, removal, and modification of architectural design decisions \cite{Garcia2013}.) In many of these cases, the only way to obtain any architectural information is to recover it from implementation-level artifacts. 
For this, a wide variety of software architecture recovery methods exists. These recover different views of a system's architecture under different paradigms. For this, they apply different algorithms on the system's implementation artifacts (e.g., the source code, bytecode, executable files, directory structure, configuration files).

Recently, a range of studies have begun looking at the nature, rate, and impact of changes in a system's architecture and the resulting architectural decay in existing systems \cite{Behnamghader2016}, \cite{Oizumi2014}, \cite{Cai2013}.
This places special emphasis on the recovery methods used in that they must be:
\begin{itemize}
\item Accurate - the architectural view they provide must be a proper reflection of the architecture.
\item Appropriately sensitive and deterministic - the difference in the obtained architectural views must be commensurate with the amount and type of system change. %
If, according to measures of architectural similarity, any change in the source code of a system results in the recovered architecture of every version of a system being entirely different from any other version, changes cannot be meaningfully compared. This diminishes the usefulness of such recoveries for evolutionary studies on the impact of changes in a system's source code on its architecture.
\item Efficient - code-bases for individual systems and different system versions must be analyzable reasonably quickly. This becomes crucial for evolutionary studies that track architectural changes over a range of versions.
\item Scalable - recovery techniques must be able to handle very large systems that are common today.
\end{itemize}

While many existing recovery methods may give an accurate view of the system under their respective paradigms, they lack one or more of the above listed desirable attributes, which limits their use and their utility in many situations. To address these issues, we have developed RELAX. We claim the architectural view (described below) produced by RELAX is useful and correctly reflects the underlying architecture. 
RELAX is appropriately sensitive to changes in that minor changes to source code do not cause major changes in the recovered view. We also claim that RELAX is both efficient and scalable, enabling it to recover architectures of large systems. This is enabled by RELAX's additivity%
, which allows the composition and reuse of partial results, distribution of the recovery process and reduction of the workload on new versions of a system to just the parts that have changed. Additionally, RELAX is tailorable by allowing different stakeholders to maximize the utility of the recovery by considering their perspective.

Just like every system has an architecture by definition (even if none of its stakeholders are aware of it), each recovery method needs to follow a paradigm that is determined by the purpose it intends to serve. We have aimed at choosing a flexible paradigm that serves as many different groups of stakeholders of a system as possible, not only one. Another goal was to make using the method and interpreting its result as straightforward as possible. It is our hope that this will lead to a democratization of architecture recovery.

These considerations have lead us to choose a concern-oriented architectural view for RELAX. In this context, a “concern” can be defined as a role, responsibility, concept, or purpose of a software system. Data persistence, Networking, and GUI are examples of generic concerns that a system may commonly address. On the other hand, there are domain-specific or application-specific concerns, e.g., Interrupt Handling as part of an OS Kernel.
(It is important to note that in the context of software architecture recovery, the noun \textit{concern} is used in its more general meaning of \textit{something that is regarded as important} and not limited to \textit{something that causes worry} \cite{Merriam-Webster}.)

Approaching a system from this point of view is useful for many different types of stakeholders: Maintainers and particularly programmers 
will be interested in learning what a system does and how and where it does it. 
A concern-centric view can also be useful for stakeholders other than programmers. For instance, the architect can assess how well concerns are separated. Project managers 
can determine task allocation among programmers with varying  degrees of familiarity with the system. Customers for whom the system is being built can check whether their concerns are reflected in it. Even interested end users may use RELAX to find out whether a system's source code implements a functionality that may not be mentioned in its documentation. The latter two types of stakeholders do not even need to be experts in software development to derive utility from RELAX.
The usefulness of a perspective based on a system's implemented concerns in comprehending an architecture has been shown \cite{Garcia2011}.  

Given an input of a system's source code and a set of concerns, RELAX classifies and clusters a system's code entities 
into word classes that relate to user-specified concerns. Its output is a view that represents the system's architectural structure and location of concerns textually and visually. Both elements of the view provide actionable information to its maintainers. Additionally, the visualization allows the viewer to gather important facts about the overall architecture of the system at a glance while also allowing them to dig deeper.

RELAX is evaluated on a set of open source systems.

The research contributions of this paper are RELAX (RELiable Architecture EXtraction), a concern based architecture recovery method that is scalable, accurate and appropriately sensitive. RELAX provides an integrated visualization of the results that can be easily interpreted and directly applied to the maintenance of the system.

The remainder of this paper is organized as follows: Section \ref{sec:foundation} explains the foundation of RELAX. Section \ref{sec:approach} describes RELAX's approach. Section \ref{sec:evaluation} presents our evaluation results. Section \ref{sec:related} compares our approach to that of other recovery methods. Section \ref{sec:conclusion} with our conclusion and section \ref{sec:future} on future work round out the paper.

\section{Foundation}\label{sec:foundation}
\subsection{Software Architecture and Architecture Recovery}
Many different definitions of ``Software Architecture'' exist \cite{SoftwareEngineeringInstitute}. Additionally, many different recovery methods exist that espouse different views of a software architecture \cite{Ducasse2009}. This creates the potential of a mismatch if both are not selected in light of each other.

Architecture recovery is the process of retrieving a system's architecture from its implementation-level artifacts \cite{Taylor2010}. Since this means that only what is actually present in the system's implementation can be used for recovery, and consequently, that the definition of ``Software Architecture'' which forms the basis of a given recovery method needs to reflect this for consistency.

This makes a definition such as ``the set of principal design decisions about a system'' \cite{Taylor2010} unsuitable for the purposes of architecture recovery, since due to erosion and drift, there is no guarantee that a single one of these decisions is realized in a current version of the system as built. In extreme cases, they may never have been present at all, rendering any attempt at recovering an architecture under this definition moot. %

A definition that fits architecture recovery in general well is: ``Fundamental concepts or properties of a system in its environment embodied in its elements, relationships, and in the principles of its design and evolution'' \cite{ISO/IEC/IEEE}. This definition %
covers the source code, which most recovery methods as well as ours are using for their basic resource, as an element of the system.
We think that another definition fits the concern-orientation of RELAX even better \cite{gacek1995definition}. According to it, a software architecture comprises 
\begin{itemize}
    \item A collection of software and system components, connections, and constraints.
    \item A collections of system stakeholders' need statements.
    \item A rationale which demonstrates that the components, connections, and constraints define a system that, if implemented, would satisfy the collection of system stakeholders' need statements.
\end{itemize}
When considering that the output of any recovery method is a view of a system's architecture, it needs to be kept in mind that there is no single ``correct'' view %
that contains the whole truth %
Instead, the same architecture can be described through different views \cite{Maqbool2007}.
While it is possible to recover the architecture of very small systems manually, for large systems with millions of SLOC only computer-aided recovery is feasible. %

\subsection{Text Classification}

\begin{figure}
	\centering
	\includegraphics[width=1.0\linewidth]{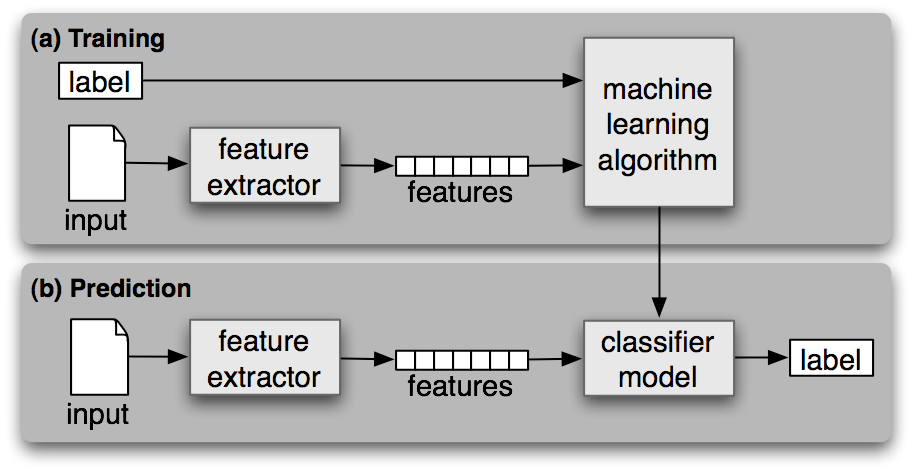}
	\caption{Text Classification Workflow}
	\label{fig:classificationworkflow}
\end{figure}

Text classification is the natural language processing method RELAX employs to locate concerns in a software system. This method automatically assigns provided documents to specified categories %
\cite{manning2008introduction}. %
To achieve this, a set of labeled examples from two or more categories of interest are supplied and then used to train a classifier algorithm which can then determine whether any given document %
belongs in one of the categories the classifier was trained on. %

Figure \ref{fig:classificationworkflow} shows the training and prediction phases of text classification and how they interact \cite{Bird2009}.

\section{Approach}\label{sec:approach}
As its input, RELAX takes a software system's source code alongside a trained classifier. It then leverages 
text classification techniques to incrementally tag individual source entities with attributes and group them into concern-related clusters. The combination of choices we made in the way in which RELAX uses text classification to build its architectural view results in a set of features %
that address several weaknesses present in other recovery methods.

\begin{enumerate}

\item The basic building block of RELAX's architectural view is the result of the
\emph{independent text classification} of each source code entity.
\item The \emph{additive nature} of forming clusters from individual source code entities whose attributes are independent of each other directly facilitates RELAX's scalability. 
\item The  \emph{Na\"ive Bayes classifier-based algorithm} further aids the scalability and accuracy of RELAX.
\item Explicit \emph{prevention of crosstalk} between changes in individual code entities 
limits their impact on the resulting architecture.
\item The \emph{ability of users to select the concerns} on which RELAX bases its recovery enables tailoring RELAX to  specific needs through easily understandable choices. 
\item An \emph{intuitive and informative visualization}  allows  stakeholders to quickly get an overview of the prevailing system-level concerns, and also to dig deeper to the level of individual source code entities. 
\end{enumerate}
This section describes the key principles underlying RELAX and its visualization, as well as the details of its implementation.
\subsection{Main Recovery Process}
\subsubsection{Selecting Concerns}

The stakeholders and their concerns stand at the beginning of the process. Those concerns can have any level of granularity, ranging from top level concerns (e.g. \textit{Database}, \textit{Graphics} or \textit{Networking}) to lower application-specific levels (e.g. \textit{HDFS Upgrade Management} or \textit{InterDataNode Protocol} for Apache Hadoop \cite{hadoop2018}). In addition, non-functional concerns (e.g. \textit{Security}, \textit{Backup}, \textit{Interoperability}) can also be used.
RELAX does not impose a hard limit on the number of input concerns to use in the training phase. %
The ``right'' number of concerns to look for in a system is not determined by any attributes of that system, such as its size or complexity. Instead, based on their knowledge about the system to be recovered and their use case, users can decide on any set of named concerns that form the basis of the system's recovery. 

For example, a project manager might be interested in a suitable task distribution of maintenance activities among programmers with specific skills. The project manager can then choose to conduct a coarse grained recovery with a selection of topics that mirrors the fields of specialization of the programmers, such as \textit{Database}, \textit{Graphics} and \textit{Networking}. In another situation, a researcher may be interested in how certain concerns are shared among related systems. For example, they could be interested in whether a project like Apache Chukwa \cite{Chukwa}, which is built on the Apache Hadoop File System (HDFS) \cite{HDFS} addresses HDFS-related concerns such as \textit{HDFS Upgrade Management} or \textit{InterDataNode Protocol}.
The choice of concerns is the only activity required of the user that is similar to setting parameters in other recovery methods. However, RELAX aims to make this an intuitive choice because the concerns are either named for well-known topics of general interest or, optionally, named by the users themselves. 

\subsubsection{Collecting Training Data and Training a Classifier}
The kind and amount of work necessary in this step depends on the concerns selected by the user. If the concerns are already covered by an existing classifier that is provided by RELAX or that the user otherwise has access to (such as through having trained it in an earlier recovery), no additional work is necessary here. 

\begin{figure}
	\centering
	\includegraphics[width=1.0\linewidth]{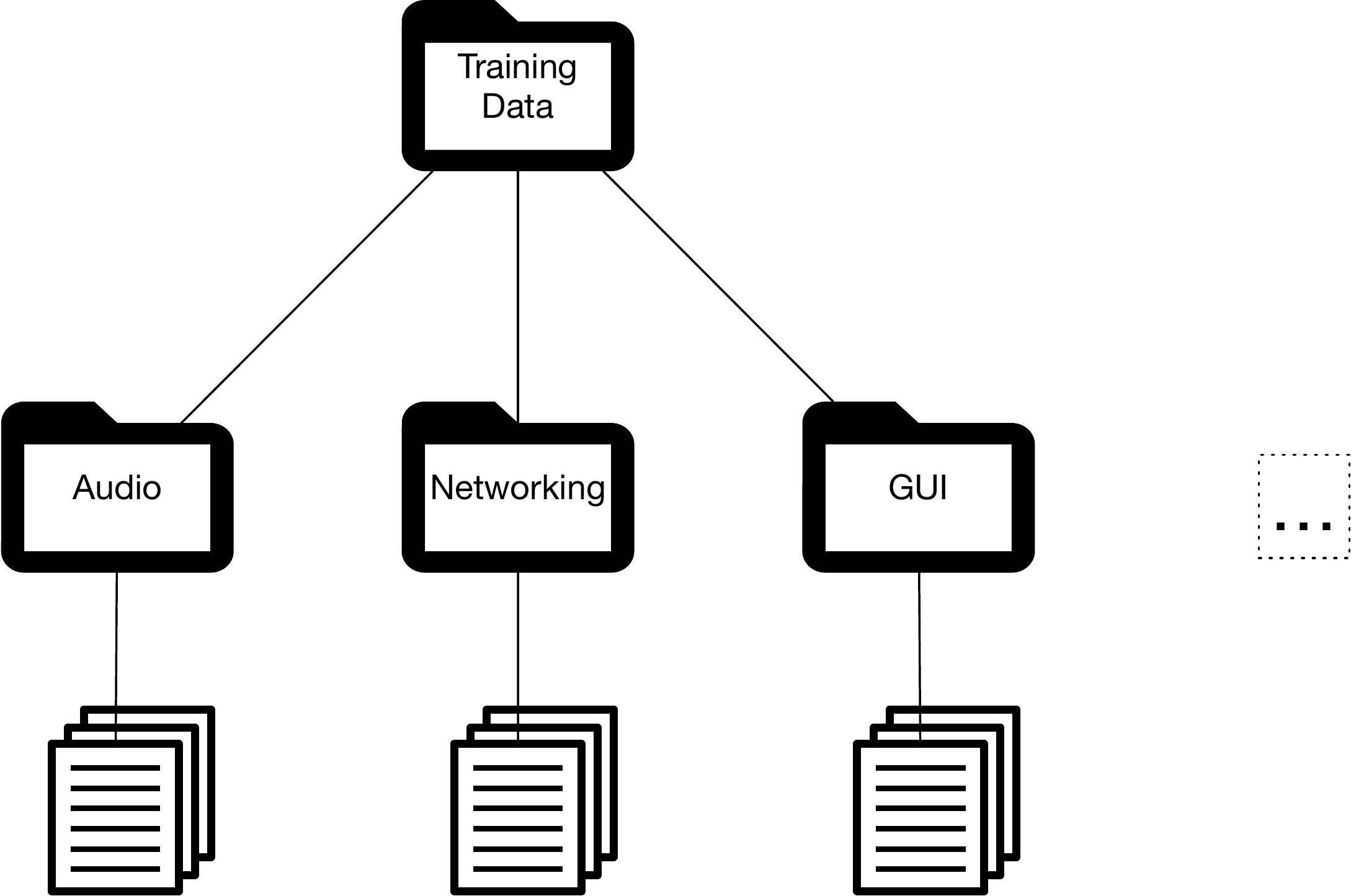}
	\caption{Training Data Example}
	\label{fig:training_data}
\end{figure}

If this is not the case, the user will be interested in training their own classifier on their chosen concerns. The required training data can either come from the curated labeled training data already provided by RELAX, or it can be provided by the user. In the latter case, the user needs to find sources of training data related to the desired concerns and label them with names of their choice. This can be any mixture of source code, API documents, articles on the subject or simply a list of related words. It is important to note that the user is not required to fully understand the training data. %
Subsequently, the user needs to label the different categories of training data with the concern names of their choosing. Figure \ref{fig:training_data} shows an example of a directory structure with training data files.

A classifier is then trained from the provided labeled training data. %
For this, the distinguishing features of the sets of documents labeled with different concerns are determined by the classifier training algorithm. 

From this, a classifier model is generated that can later be used to label different sets of data that were not part of the training process, such as the code entities of systems whose architecture is to be recovered.

\begin{figure*}[t]
	\centering
	\includegraphics[width=1.0\linewidth]{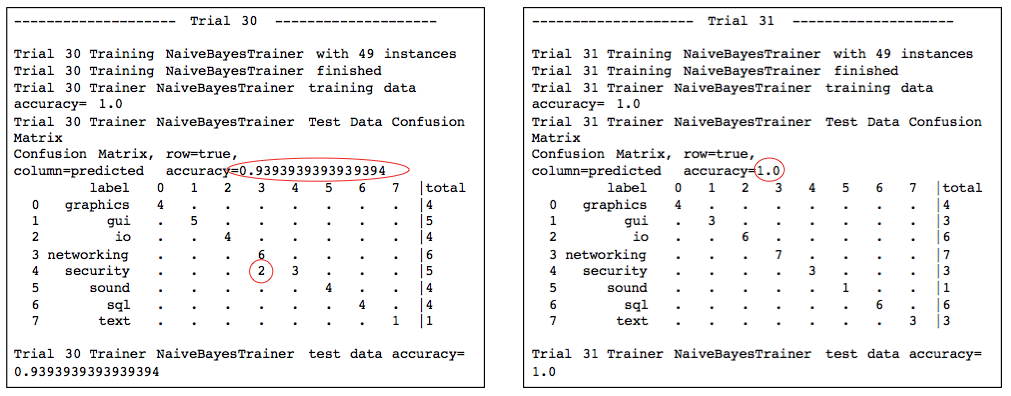}
	\caption{Classifier Candidate Selection}
	\label{fig:clscandsel}
\end{figure*}

The training process generates a number of classifier candidates whose accuracy is then checked on a portion of the training data that has been set aside for this purpose. The classifier with the best accuracy is then chosen as the classifier to be used for architecture recoveries. Figure \ref{fig:clscandsel} shows the accuracy information obtained for two classifier candidates called ``Trial 30'' and ``Trial 31''. It shows the overall accuracy of each classifier candidate as well as a confusion matrix. A confusion matrix is a table whose rows show the labels of the test data and whose columns show the labels determined by the classifier candidate. It is easy to determine whether a candidate's results are fully correct (i.e., all documents are labeled by the classifier candidate with the labels of the test data) by looking at whether or not all numbers are lined up on the diagonal from the table's origin to the lower right. As a summary, an overall accuracy value, between 0 and 1, for a candidate can be computed by dividing the number of correctly identified test documents by the overall number of test documents. The training output also shows the overall accuracy value for the candidate as a value between 0 and 1 that is calculated by dividing the number of correctly identified test documents by the overall number of test documents.
In the case of ``Trial 30'' in the figure, we can see that  While ``Trial 30'' has misclassified two test documents that should have been labeled with ``security'' as ``networking'' and therefore has an accuracy close to 0.94, or 31/33. ``Trial 31'' has classified all documents correctly and consequently has an accuracy of 1.0. It should therefore be chosen.

Once a classifier is trained on a set of topics, it is reusable.

\subsubsection{Classification}
\hfill

\lstset{language=Java}          %
\begin{lstlisting}[basicstyle=\tiny, float,frame=single,style=base,caption={Example code with SQL-related text in red},captionpos=b,label={lst:codesnippet}]  %

package org.apache.hadoop.chukwa.@database@;
import org.apache.hadoop.chukwa.util.@Database@Writer;
import java.@sql.SQLException@;
import java.@sql.ResultSet@;
import java.@sql.ResultSetMetaData@;

public class Consolidator extends Thread {
  private @Database@Config dbc = new @Database@Config();
  private String @table@ = null;
  public Consolidator(String @table@, String intervalString) {
    super(@table@);
        this.@table@ = @table@;
    }}
  String @query@ = ``@select@ * from '' + @table@;
  log.debug(``@Query@: '' + @query@);
  rs = db.@query@(@query@);
  if (rs.next()) {
@ResultSetMetaData@ rmeta = rs.@getMetaData@();
    for (int i = 1; i <= rmeta.@getColumnCount@(); i++) {
      if (rmeta.@getColumnName@(i).equals(``timestamp'')) {
        start = rs.getTimestamp(i).getTime();
        }
      end = start + (interval * 60000);
        } catch (@SQLException@ ex2) {
      log.error(``Unable to determine starting point in @table@: '' + this.@table@);
      log.error(``@SQL@ Error:'');
return;
}
\end{lstlisting}
In the classification step, the trained classifier extracts the features from each code entity and assigns a feature vector to it based on that entity's affinities to each concern the classifier has been trained on.
We have chosen Na\"ive Bayes as a classifier for RELAX based on several considerations: first, it assumes that features are independent, which appears to be a good fit for code files, where each feature may be encountered individually and can individually determine which topic a code entity belongs to. Second, its linear time complexity serves the scalability of RELAX. Third, classifiers trained with it have performed well in our accuracy evaluation (compare to Section \ref{sub:Accuracy}). %
Last but not least, the prediction model of the Na\"ive Bayes algorithm is deterministic \cite{Krzysztofowicz1999}. Determinism is an important feature for evolutionary software studies since without it, we cannot determine with certainty whether two different recovered architectural views which were produced by the same recovery method came from two different systems or system versions. 
Listing \ref{lst:codesnippet} shows a database-related code snippet with the words that indicate its relation to SQL databases highlighted. For a Na\"ive Bayes classifier, the feature vector consists of values between 0 and 1 for each concern. For example, the feature vectors for three code entities called ``SQL.Java'', ``Screen.Java'' and ``ConnectIP.Java'' could look like the rows of Table \ref{tab:entities}. We can see that the affinity values over all concerns do not have to add up to 1.0 and that they can have values that are not 0 or 1. This is because a code entity may not be related to any selected concern or it may be strongly related to more than one concern.
 
\begin{table}[bp]
\renewcommand{\arraystretch}{1.3}
\caption{Entities with Feature Vectors}
\centering
\rowcolors{2}{white}{gray!25}
\begin{tabular}{ |c| c c c|}
\hline
 Entities & Database & Graphics & Networking \\ 
 \hline
 SQL.Java & 0.9 & 0.1 & 0.2\\  
 \hline
 Screen.Java & 0.05 & 0.95 & 0.1\\
 \hline
 ConnectIP.Java & 0.02 & 0.01 & 0.92\\
\hline
\end{tabular}
\vspace{10px}
\label{tab:entities}
\end{table}

\subsubsection{Clustering}

\begin{table}[bp]
\renewcommand{\arraystretch}{1.3}
\caption{Clusters with Feature Vectors}
\centering
\rowcolors{2}{white}{gray!25}
\begin{tabular}{ |l| c c c|}
\hline
\diagbox{Cluster}{Feature} & Database & Graphics & Networking \\ 
 \hline
 Database & 1 & 0 & 0\\  
 \hline
 Graphics & 0 & 1 & 0\\
 \hline
 Networking & 0 & 0 & 1\\
 \hline
 Unknown & 0 & 0 & 0\\
 \hline
\end{tabular}
\vspace{10px}

\label{tab:clusters}
\end{table}
Before clustering begins, each user-selected concern-related cluster is assigned an orthogonal feature vector that mirrors that concern and allows code entities to be grouped into it. A default ``Unknown'' cluster without any concern affinities is always created for the code entities that are not related to any selected concern. The rows in Table \ref{tab:clusters} show the feature vectors for three clusters related to databases, graphics and networking, respectively as well as the default cluster for entities that are not related to any selected concern.
Based on the results of the classification, each code entity is then assigned to the concern-related cluster that its feature vector is most similar to. This similarity is determined using the cosine similarity between the feature vector of the code entity and the cluster. Cosine similarity is a measure of the distance between vectors and is commonly used in Natural Language Processing in order to determine how close a body of text is to a given topic \cite{Baker1998}, \cite{Huang2008}, \cite{Lin2014}.

\subsubsection{Additivity and Crosstalk Prevention}

Recall that our goals for RELAX include scalability, efficiency, appropriate sensitivity and determinism. Our intuitive approach to this is to explore building up the overall recovery result from individual parts that could be individually and independently processed and reused or updated as needed. We then decided that these individual parts should be the source code entities of the system and analyzed which beneficial properties would emerge.

RELAX classifies each code entity as belonging to a set of user defined concerns. The classification task is performed on an individual source code entity and has no dependence on the classification of any other entity.

The classification of source code entities individually enables RELAX's important property of additivity. This means that the recovery results of the whole system or its subsystems can be composed from smaller parts, eventually reaching down to the ground level of individual source code entities. Additivity in turn enables scalability and efficiency by allowing the following operations:
\begin{itemize}
\item For the architecture of a system to be recovered, it can be split up into smaller units which can then be distributed to be classified and associated with a cluster. This way, the ceiling of the system size that can be evaluated is nearly unlimited.
\item For evolutionary studies on a system, only the entities
that have changed will need to be evaluated. The information on the remaining entities gained in a previous recovery run can be reused.
\item Libraries or frameworks can be evaluated separately and their results added or subtracted from the whole as needed.
\end{itemize}

Further, the individual classification also limits ``Crosstalk''. Crosstalk is a phenomenon in which a change in a source code entity affects parts of the recovered architectural view that do not pertain to it. Therefore, without Crosstalk affecting RELAX, it means that the change of the recovery result is only confined to the code entities that have changed and (possibly, depending on whether their associated concerns have changed) their cluster association. Further, the scale of the changes in the recovered view is proportional to with the size of the changes in the source code entities.

\subsubsection{Textual Output}

Conceptually, the textual output produced contains (1) The classification of each source code entity, (2) the constituents of all concern-related clusters, and (3) the auxiliary output from other tools, such as the list of dependencies between code entities or the size of entities in SLOC, which can be used for further processing and analysis. (RELAX uses the Classycle library \cite{Elmer2014} to determine the dependencies between code entities. are not determined by RELAX, but can be used for further analysis.)

\subsection{Visualization} 
\begin{figure}
	\centering
	\includegraphics[width=1.0\linewidth,trim=8 6 10 6,clip]{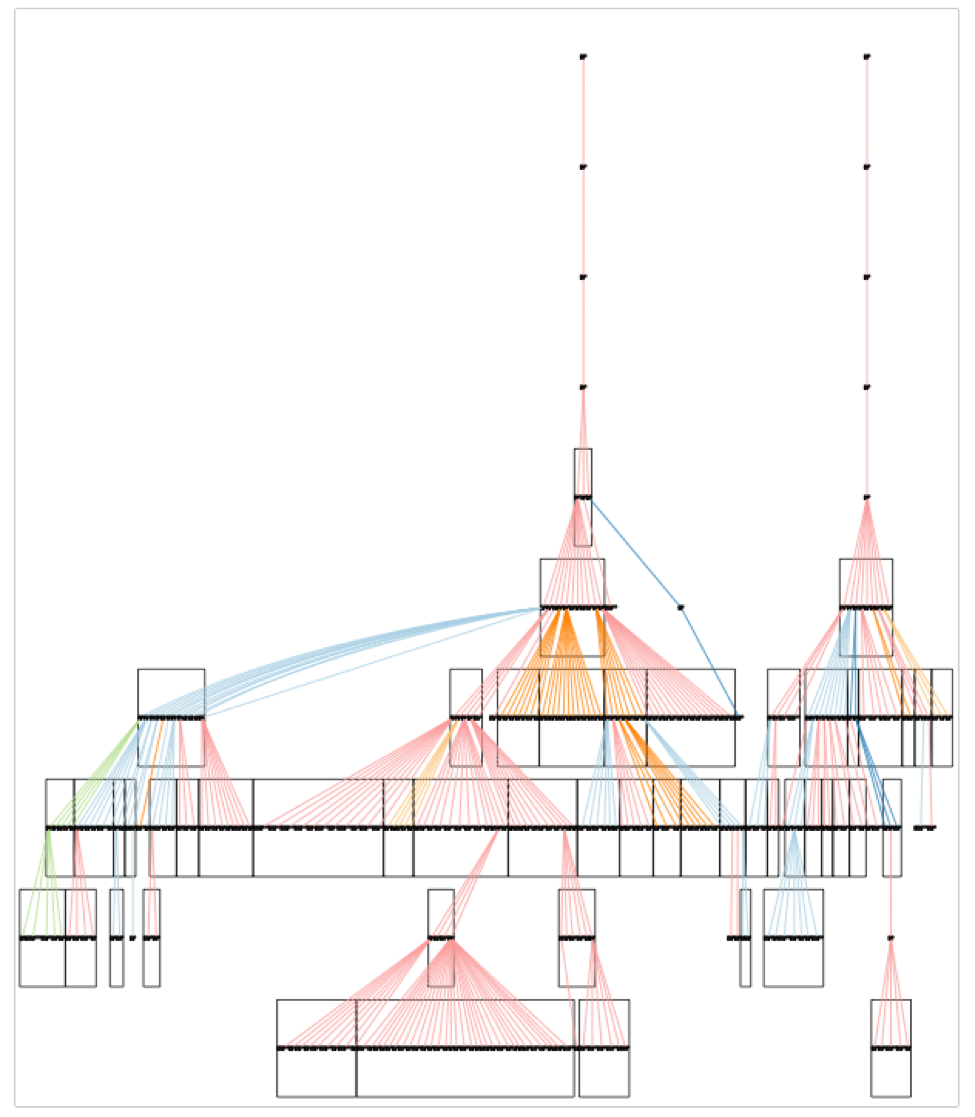}
	\caption{RELAX Directory Graph Example}
	\label{fig:dirgraphex}
\end{figure}

The directory visualization of RELAX shown in Figure \ref{fig:dirgraphex} aims to give any stakeholder a high-level overview of the system architecture and the system's addressed concerns that can be enlarged to the level of individual source entities.

The visualization is based on the directory structure of the system, which corresponds to the package structure in Java. The system is shown as a directory tree. Nodes are either packages (inner nodes) or source entities (leaf nodes).
Nodes that belong to the same package are surrounded by a rectangle. Since software systems can consist of a very large number of source entities, individual nodes can be very small in an overview (situations in which an individual node would make up less space than a pixel would be conceivable), and gaining an impression of their concerns would be impossible. 

 Therefore, in order to guarantee that concerns can be shown, the lines from each package folder to its children are shown in the color prevailing that corresponds to the prevailing concern in that package. 
The prevailing concern is determined as follows: For each child node, the main concern is determined by the classifier as the topic most relevant to the corresponding code entity. The weight of this entity is then determined by its file size (physical or logical SLOC can also be selected for this). If a child node is not a leaf node (i.e. it stands for a package), then its prevailing concern is the concern that carries the most weight with its children. This relationship holds recursively throughout the tree.
One important outcome of this is that there is an easy way to see what the main concern of the overall system is by checking the color of the root node (or colors of the root nodes, if several exist) of the system. Because of the recursion, this holds for each package.

RELAX generates a legend for the directory visualization which shows the names of the concerns as they exist in the classifier in the color automatically selected for them by RELAX. The color selection is based on guidelines for optimal distinguishability of adjacent colors \cite{Brewer}.

\begin{figure}
	\centering
	\includegraphics[width=1.0\linewidth]{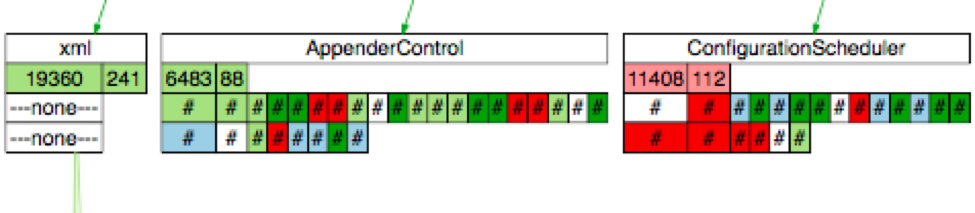}
	\caption{RELAX Directory Graph Detail}
	\label{fig:dirgraphdet}
\end{figure}

Individual nodes can be examined by zooming in. The paradigm that this  visualization is following is that of a navigational file manager, such as the Finder in macOS or the Explorer in Windows. The details shown in Figure \ref{fig:dirgraphdet} correspond to the metadata view obtained by right-clicking and selecting ``Get Info'' on the macOS Finder or right-clicking and selecting ``Properties'' on the Windows Explorer. In the example shown in Figure \ref{fig:dirgraphdet}, we are seeing a package (left) and two Java source entities. Since all three belong to the same package, we see three incoming arrows from the top in the same color %
Each entity is shown with a group of attributes:
\begin{itemize}
	\item A top box containing the base-name of its canonical name,
	\item A second row showing
	\begin{itemize}
		\item its file size in bytes with the color of its concern as the background color,
		\item its logical SLOC with the same background color.
	\end{itemize} 
	\item A third row with all outgoing dependencies colored for the corresponding entities,
	\item A fourth row with all incoming dependencies colored for the corresponding entities.
\end{itemize}

Checking individual entities can give the user an impression of how connected an entity is and which type of concerns the related entities address. Questionable dependencies could be caught here.
The format of the file that is used to lay out the directory graph is a human-readable text file that describes a directed graph. The actual layout is done by dot, a program from the Graphviz \cite{AT&TResearch} package. The dot program creates hierarchical layouts. Results are created in PDF format. %
It is possible to provide specific directives for the width and height of the graph. %

From the hierarchical diagram of the system shown in Figure \ref{fig:dirgraphex}, a stakeholder can immediately get an overview of the system and gain some first impressions: First, it is apparent that the system has two top level folders (with branching only beginning several levels below the top due to the Java packaging conventions, which use the reverse Internet domain names of organizations \cite{Oracle2015}). It is clear that five package levels have leaf nodes (which stand for code entities). The third level from the bottom has the most code entities. 

Regarding concerns, the system seems to be mostly addressing the one that is shown in bright red. Two concerns, bright green and dark blue, seem to be addressed mostly in one package each (second level from the bottom at the very left and near the middle of the third level from the bottom).
Several concerns, such as the orange, the light blue one and chiefly the bright red one, are shown to be distributed throughout the system. This could indicate a poor separation of concerns (or possibly the need for a narrower definition of the concerns that should be used for classification).

Conclusions can also be drawn when studying the evolution of a system. The diagrams in Figures \ref{fig:dirgraphfirst} and \ref{fig:dirgraphsecond} show two consecutive minor versions of the same system:

\begin{figure}
	\centering
	\includegraphics[width=1.0\linewidth]{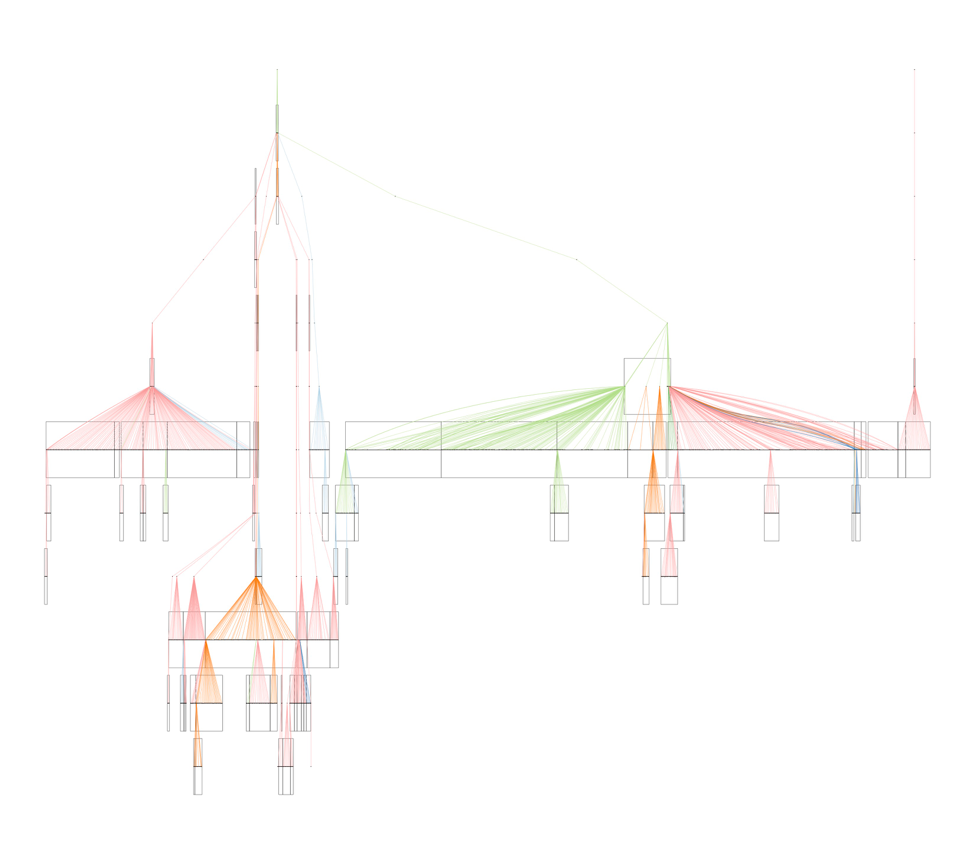}
	\caption{RELAX Directory Graph of First System Version}
	\label{fig:dirgraphfirst}
	\includegraphics[width=1.0\linewidth]{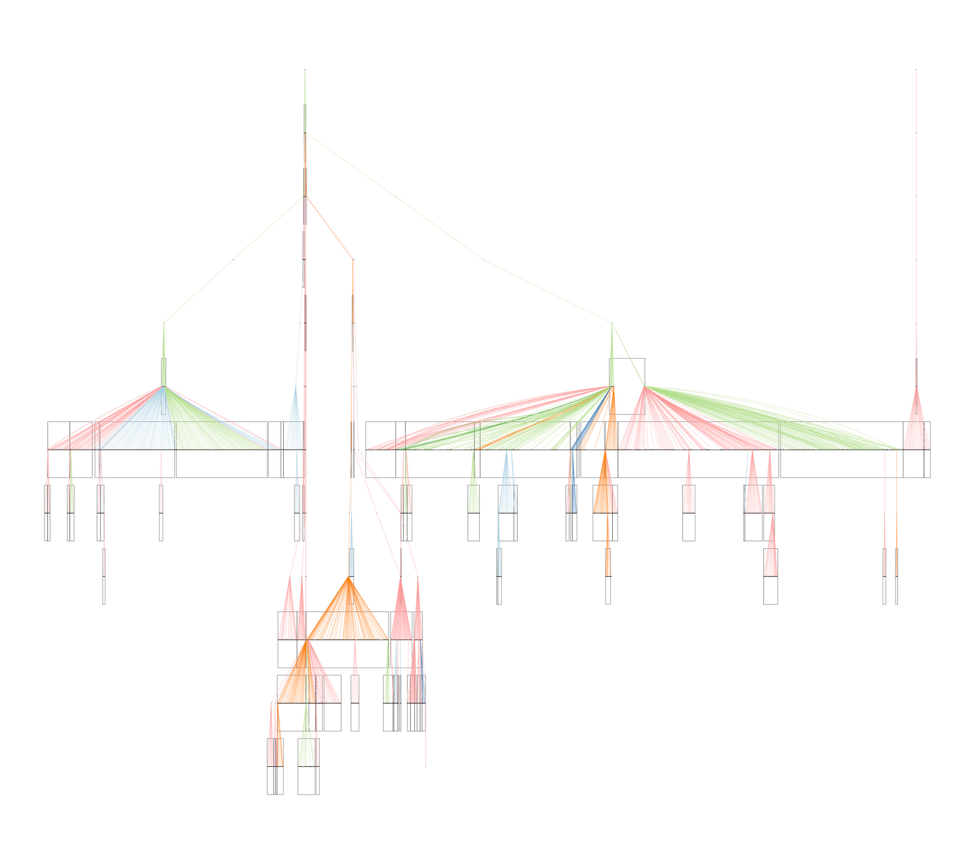}
	\caption{RELAX Directory Graph of Second System Version}
	\label{fig:dirgraphsecond}
\end{figure}

The similar outlines are making the two versions of the system easy to compare (though some differences in shape are due to the automatic layout in the Graphviz package). The comparison shows that the leftmost package in the hierarchy, which was dominated by the red concern in the first version is now more evenly split three ways between red, blue and green and has changed its prevailing concern from red to green.

\subsection{Workflow}

\begin{figure}
	\includegraphics[width=1.0\linewidth]{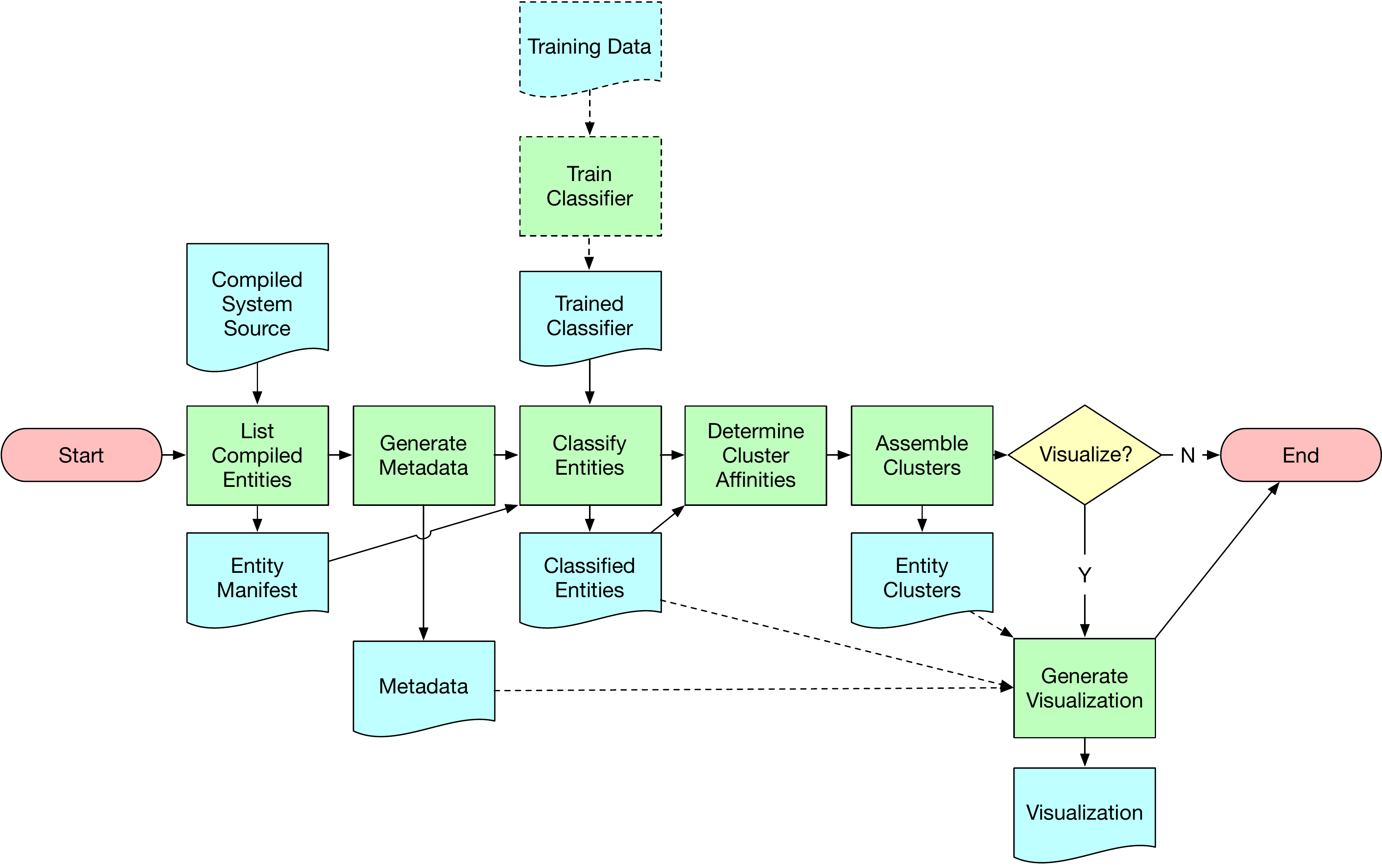}
	\caption{RELAX Recovery Workflow}
	\label{fig:recworkflow}
\end{figure}

Figure \ref{fig:recworkflow} shows the workflow of a RELAX recovery from the point of view of the programmatic process, which incorporates all parts of our approach. The selection of concerns is not shown as an explicit step, but is an implicit part of the selection of a trained classifier. Training a classifier based on training data is shown with dashed lines since it is not a necessary part of each recovery.

\subsection{Implementation}
RELAX uses the MALLET \cite{McCallum} toolkit, which includes different classification algorithms such as  Na\"{\i}ve Bayes \cite{Murphy2006}\cite{Li1998}, Maximum Entropy \cite{Nigam1999} and Decision Trees \cite{Safavian1990} and allows training and applying them.
 
RELAX has been implemented in Java as part of a workbench comprising of a suite of architecture recovery techniques. The implementation is GUI-based and allows training classifiers, running RELAX, and visualization of the results without leaving the GUI. The principal output produced is a textual clustering of the system's source code entities and a directory visualization. %
\section{Evaluation}\label{sec:evaluation}

\subsection{Accuracy}\label{sub:Accuracy}

Some of the principal results of a RELAX recovery are the classification of all individual code entities of a system and their grouping into a set of concern-related clusters.
The accuracy of the clustering can be determined by measuring its similarity to an expert decomposition, which is another clustering manually prepared by an expert on the system, such as its architect. 
The expert decomposition serves as a ``ground truth'' \cite{Maqbool2004a}. A known measure of similarity for this is MojoFM \cite{ZhihuaWen2004}. We picked it for our evaluation because it has been used in several studies such as \cite{patel2009software} as well as \cite{mahmoud2013evaluating} and data for our evaluation, which compares the  respective closeness of RELAX to expert decompositions is already available for ACDC and ARC (see Section \ref{sec:related}), while the clustering results that formed the basis of the study \cite{Garcia2013} are not. We felt it was important to compare the performance of RELAX to that of ARC, since the latter is another recovery method whose paradigm is that of concern-oriented clustering.

It expresses the similarity between two partitions of a set as a percentage, where 100\% represents identity and 0\% maximal difference. Its formula is:
\begin{equation}
MojoFM (M) = (1-\frac{mno(A,B)}{max(mno(\forall A,B))})\times100\%
\end{equation}
Where $M$ stands for the clustering technique. $A$ is the clustering produced by $M$ and $B$ is the expert decomposition. $mno(A,B)$ represents the minimum number of \emph{Move} (moving an object to a different cluster) and \emph{Join} (joining one or more clusters to form a new cluster) operations to transform partition A to B.

For the purposes of our comparisons of RELAX clusterings to expert decompositions, we are interested in answering the following question: How close would RELAX come to the expert decomposition if a classifier would be trained to categorize a system's code entities into clusters related to the concerns present in the expert decomposition?

Table \ref{tab:mojofmvalues} compares the MojoFM values of RELAX to those of two other recovery methods (ACDC and ARC, both of which are described in detail in Section \ref{sec:related}) that have been identified previously as the two closest to the expert decompositions of eight systems out of a set of ten recovery methods \cite{Garcia2013}. As can be seen in the diagram, RELAX exceeds their MojoFM values in five cases, is between the two in one case and closely below them in two. This lets us conclude that RELAX's overall accuracy is better than that of the two most accurate known recovery methods so far.

\begin{center}
	\begin{table}[tp]
		\caption{MojoFM Ground Truth Comparison Values}
		\renewcommand{\arraystretch}{1.3}
		{\small
			\hfill{}
			\rowcolors{2}{white}{gray!25}
			\begin{tabular}{|r|r|r|r|}
				\hline
				\textbf{System}&\textbf{RELAX}&\textbf{ARC}&\textbf{ACDC}\\
				\hline
				Bash&\adjustbox{margin=1ex,bgcolor=orange}{75.86}&\adjustbox{margin=1ex,bgcolor=gray!25}{57.89}&\adjustbox{margin=1ex,bgcolor=gray!25}{49.35}\\
				OODT&\adjustbox{margin=1ex}{43.47}&\adjustbox{margin=1ex,bgcolor=orange}{48.48}&\adjustbox{margin=1ex}{46.01}\\
				Hadoop&\adjustbox{margin=1ex,bgcolor=gray!25}{58.32}&\adjustbox{margin=1ex,bgcolor=gray!25}{54.28}&\adjustbox{margin=1ex,bgcolor=orange}{62.92}\\
				ArchStudio&\adjustbox{margin=1ex}{74.60}&\adjustbox{margin=1ex}{76.28}&\adjustbox{margin=1ex,bgcolor=orange}{87.68}\\
				Linux-D&\adjustbox{margin=1ex,bgcolor=orange}{74.67}&\adjustbox{margin=1ex,bgcolor=gray!25}{51.47}&\adjustbox{margin=1ex,bgcolor=gray!25}{36.31}\\
				Linux-C&\adjustbox{margin=1ex,bgcolor=orange}{93.70}&\adjustbox{margin=1ex}{75.72}&\adjustbox{margin=1ex}{63.76}\\
				Mozilla-D&\adjustbox{margin=1ex,bgcolor=orange}{53.47}&\adjustbox{margin=1ex,bgcolor=gray!25}{43.44}&\adjustbox{margin=1ex,bgcolor=gray!25}{41.20}\\
				Mozilla-C&\adjustbox{margin=1ex,bgcolor=orange}{90.62}&\adjustbox{margin=1ex}{62.50}&\adjustbox{margin=1ex}{60.30}\\
                \textbf {Average}&\adjustbox{margin=1ex,bgcolor=orange}{70.59}&\adjustbox{margin=1ex,bgcolor=gray!25}{58.76}&\adjustbox{margin=1ex,bgcolor=gray!25}{55.94}\\
				\hline
		\end{tabular}}
		\hfill{}
		\vspace{10px}
	
		\label{tab:mojofmvalues}
	\end{table}
\end{center}
\subsection{Scalability and Efficiency}
Since each file is classified individually and independent of any other, and the time of an individual Na\"ive Bayes classification depends only on the size of the file to be classified \cite{Lewis1994}, the time required to recover a system's architecture should scale linearly with the overall size of files in a system to be classified, or the lines to be processed.

In order to determine how the performance of RELAX changes with the system size, its performance was measured with versions 15 versions of Apache Hadoop \cite{hadoop2018}, 7 versions of Apache Chukwa \cite{Chukwa}, and one version each of Log4j2 \cite{log4j2} and Chromium \cite{chromium}. Altogether, these comprised more than 2.45 million SLOC.

\begin{figure}
	\centering
	\includegraphics[width=1.0\linewidth,trim=4 4 4 4,clip]{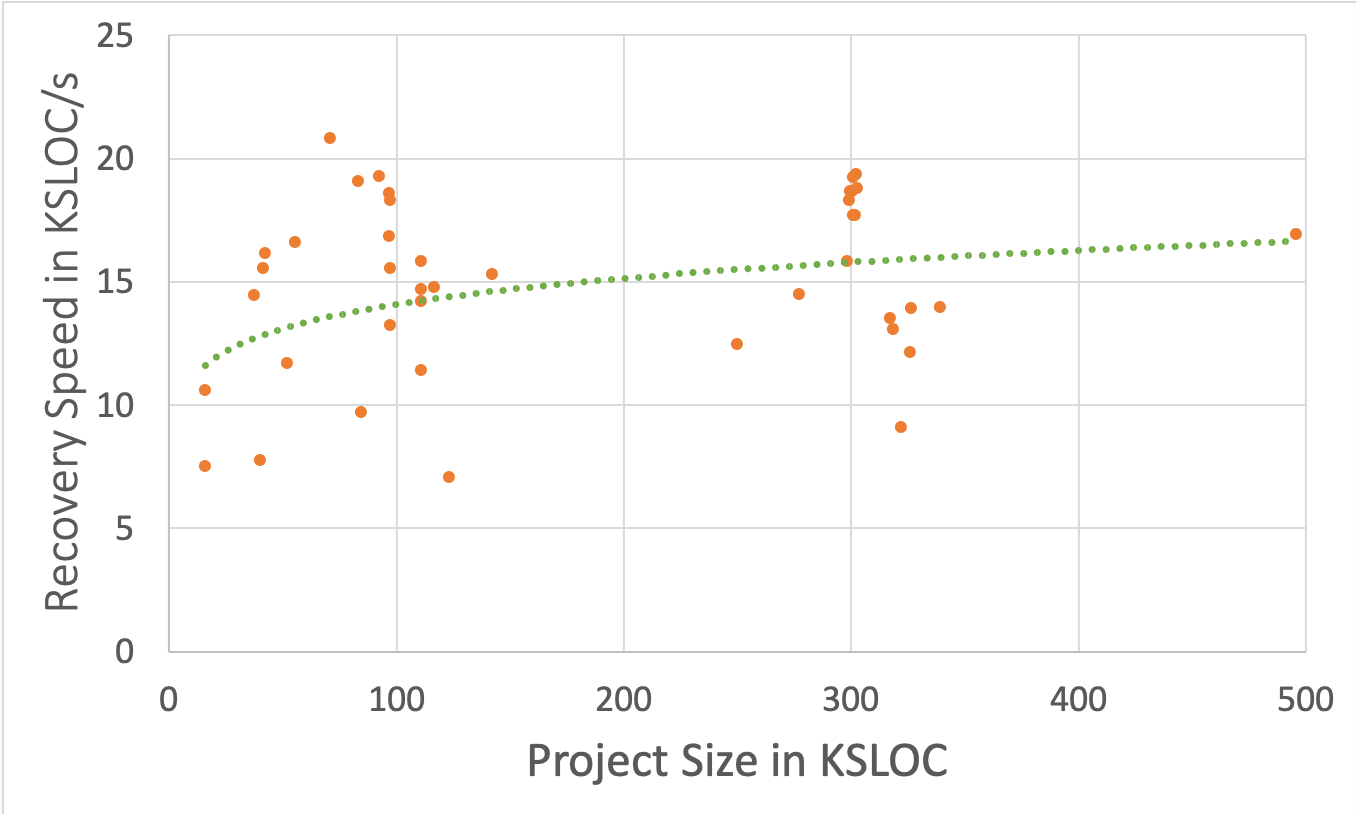}
	\caption{RELAX Recovery Speeds for Projects of Different Sizes}
	\label{fig:physical_SLOC}
\end{figure}

The scatter plot in Figure \ref{fig:physical_SLOC} shows our observations of how many SLOC were processed per second by RELAX, respectively, for each of the 24 systems. A trend line was fitted to the plot.

The observations confirm that, as expected, the number of source lines of code (SLOC) that are processed over a given unit of time does not decline with the size of the system. (The trend line shows an increasing performance with an increase in SLOC. This may be an artifact of the underlying OS and is not expected to be sustained for bigger system sizes.) %
\section{Related Work}\label{sec:related}
\subsection{PKG}
PKG \cite{Le2015}, the simplest approach to architecture recovery is based on the package-level structure view of a system's implementation. This approach produces an objective but not architecturally satisfying view in that it stays at the surface instead of trying to assist its user to determine why the system is built the way it is. %
Other clustering techniques have been suggested on the basis on file names and file naming conventions \cite{Anquetil1997a, Anquetil1999a}. However, their assumptions about naming conventions are not always correct. Many other sophisticated techniques exist. %
We will review several relevant approaches.%

\subsection{ACDC}

The ACDC (Algorithm for Comprehension-Driven Clustering) algorithm \cite{tzerpos2000acdc} uses structural relationships specified as patterns to create an algorithm for recovering components and configurations that bounds the size of the cluster (the number of software entities in the cluster), and provides a name for the cluster based on the names of files in the cluster. ACDC's view is oriented toward components that are based on structural patterns (e.g., a component consisting of entities that together form a particular subgraph).

\subsection{ARC}\label{sub:arc}

ARC \cite{Garcia2011} %
uses topic modeling to find concerns and combines them with the structural information to automatically identify components and connectors \cite{Garcia2011}. The topic model employed in ARC is %
Latent Dirichlet Allocation (LDA) \cite{Blei2003a}. Using LDA, ARC can detect concerns in individual code entities and compute similarities between them.
A software system's implementation entities, such as its source files, are represented as a set of documents (a corpus) and each document in turn as a ''bag of words'' \cite{Garcia2011}. Each document can be related to several different topics. Based on those topics, the documents %
are clustered using dependencies between them as structural information %
and concerns (the topics from the topic model) as features.
It is very important to note that topic modeling as applied in ARC will not name the detected topics automatically and is an iterative process which is not guaranteed to yield topics that are consistent or that a human being can name \cite{link2019value}. In contrast, document classification uses named topics from the outset.
We have outlined issues with ARC in detail elsewhere \cite{link2019value} and will therefore limit ourselves to a short overview. They comprise
\begin{itemize}
    \item Its handling of stop words,
    \item The selection of the number of topics to be detected,
    \item Topic quality,
    \item Determinism,
    \item Sensitivity to architectural change, and
    \item Scalability.
\end{itemize}

\begin{comment}

\section{Other Related Work}

\subsection{ARC}\label{sub:arc}

\subsubsection{Stop Words}

Stop words are words that are common to any topic and whose presence therefore cannot be used to distinguish one topic from another. Some examples of stop words are ``And'',``Of'', and ``This''. These words can matter when training classifiers \cite{Chaovalit2005}, but not when running the classification. This is because the stop words can be removed from the training data and once removed, the classifier will not look for them in the bodies of text to be classified. They will therefore be ignored. %

\subsubsection{Number of Topics}

Since in text classification, topics stand at the beginning and not the end of the process, finding the correct number of topics is not an issue. Rather, that number is provided by the user and mirrors what the user is interested in.

\subsubsection{Topic Quality}

The quality of the topics is fully controlled by the user since it is the user who can define topics and train classifiers (or use the provided ones if they user is satisfied with their quality).
\end{comment}

\section{Threats to validity}\label{sec:threats}
\subsection{Generalizability}
In order for RELAX (or any other recovery method that is based on natural language processing) to be able to produce meaningful results through natural language processing, all of the following conditions need to hold:
\begin{enumerate}
\item The programming language the system is written in allows the use of comments or variable names,
\item The code contains meaningful comments or variable names,
\item The comments and variable names are pertinent to the purpose of the code.
\end{enumerate}
This excludes code that has misleading comments or variable names or has been obfuscated.
Additionally, due to availability issues regarding closed-source systems' source code, only versions of open-source systems have been evaluated.
\subsection{User Studies}
Because the selection of systems to be evaluated had not settled yet when our evaluation was conducted, user studies with engineers that would serve to further ascertain of RELAX have not been conducted yet. %
\section{Conclusion}\label{sec:conclusion}
In this paper we presented a novel architecture recovery method which employs text classification to recover a concern-oriented view of a system.
The approach classifies source code entities to clusters based on user-defined concerns. The conceptual and design choices made have resulted in an accurate and scalable concern based recovery method. The tool that implements the approach has been evaluated for accuracy and scalability on a set of open source systems. The results confirm the claims made. %

\section{Future Work}\label{sec:future}
Currently, RELAX assigns each source code entity to the cluster that represents its dominant concern. We aim to allow users to control the way in which code entities are assigned to clusters. For instance, the users could choose clusters which represent what to them are meaningful combinations of more than one concern and instruct RELAX to assign code entities to such clusters.
Another feature to be implemented is the ability to define undesirable dependencies in a system (e.g., those that break a desired layered architectures by having entities that serve low-level concerns depend on others in a higher layer.)
Of further interest are studies of architecture evolution with RELAX, as well as comparisons between its performance and that of other recovery methods on large or very large systems (e.g. Chromium OS). %
\ifCLASSOPTIONcompsoc
  \section*{Acknowledgments}
\else
  \section*{Acknowledgment}
\fi

The authors would like to thank Nenad Medvidovic for reviewing early drafts of this paper. We would further like to thank Duc Le and Suhrid Karthik for providing comments and insights. 
\Urlmuskip=0mu plus 1mu
%\bibliography{IEEEabrv,relax-ref.bib}

\end{document}